\documentstyle[aps,preprint,tighten,floats,epsf,rotate]{revtex}

\begin{document}
\draft
%%%%%%%%%%%%%%%%%%%%%%%%%%%%%%%%%%%%%%%%%%%%%%%%%%%%%%%%%%%%%%%%%%%%%%%%%

%\preprint{\vbox{\it 
%                        \null\hfill\rm    IP-BBSR/2001-30, Oct. 2001}\\\\}
%%%%%%%%%%%%%%%%%%%%%%%%%%%%%%%%%%%%%%%%%%%%%%%%%%%%%%%%%%%%%%%%%%%%%%%%%
%
\title{Sustaining supercooled mixed phase via resonant oscillations
of the order parameter}
\author{Rajarshi Ray \footnote{email: rajarshi@iopb.res.in}, Soma Sanyal 
\footnote{email: sanyal@iopb.res.in}, and Ajit M. Srivastava 
\footnote{email: ajit@iopb.res.in}}
\address{Institute of Physics, Sachivalaya Marg, Bhubaneswar 751005, 
India}
%
%\date{July 2002}
%
\maketitle
\widetext
\parshape=1 0.75in 5.5in
\begin{abstract}
 We investigate the dynamics of a first order transition when the order 
parameter field undergoes resonant oscillations, driven by a periodically 
varying parameter of the free energy. This parameter could be a
background oscillating field as in models of pre-heating after
inflation. In the context of condensed matter systems, it could be
temperature $T$, or pressure, external electric/magnetic field etc. 
We show that with suitable driving frequency and amplitude, the system 
remains in a type of mixed phase, without ever completing  transition 
to the stable phase, even when the oscillating parameter of the free
energy remains below the corresponding critical value (for example, 
with oscillating temperature, $T$ always remains below the 
critical temperature $T_c$). This phenomenon may have important 
implications. In cosmology, it will imply prolonged mixed phase 
in a first order transition  due to coupling with
background oscillating fields. In condensed matter systems, it will
imply that using oscillating temperature (or, more appropriately, 
pressure waves) one may be able to sustain liquids in a mixed phase 
indefinitely at low temperatures, without making transition to 
the frozen phase. 
\end{abstract}
\vskip 0.125 in
\parshape=1 -.75in 5.5in
\pacs{PACS numbers: 64.70.Dv, 82.60.Nh, 98.80.Cq}
Key words: {Resonance, supercooling, phase transitions, preheating, inflation} 
%\newpage
%\begin{multicols}{2}
\narrowtext
%%%%%%%%%%%%%%%%%%%

\section{Introduction}

   Recently, it has been shown in ref.\cite{rsntd} that if an order
parameter field undergoes resonant oscillations due to some parameter
of the free energy, e.g. temperature (or pressure, background field etc.),
being driven periodically, then a finite density of topological defects
can arise even when the system never goes through the phase transition,
and remains at temperatures {\it much below} the critical 
temperature $T_c$. Also, the spatial distribution of order parameter at 
late times resembled more like a system undergoing phase transition, 
rather than the ordered phase at low temperatures. It was speculated in 
\cite{rsntd} that this phase may have interesting properties. In the 
present paper we address this issue for the case of a first order
transition. We will briefly comment on the case of a second order 
transition (which was the case in ref.\cite{rsntd}), a detailed 
discussion of that case is postponed for a future publication. 

 Consider a system, supercooled, and trapped in the metastable high 
temperature phase, as denoted by the free energy plot (solid curve) in 
Fig.1. If the system is left in this stage, critical bubbles of the low 
temperature stable phase will nucleate via thermal fluctuations, will
grow, coalesce, and convert the 
entire system to the stable phase. We will study the dynamics of
this transition, using numerical simulation, when some parameter of
the free energy is undergoing periodic
variations, leading to periodic changes in the shape of the free energy
plot as indicated by dashed curves in Fig.1. 
This will drive the order parameter field away from the stable phase,
inducing field oscillations.  Occasionally, localized resonances will 
force the field to go across  the barrier, all the way up to the 
metastable vacuum. Such a region
will shrink again, while more regions bounce to the metastable phase.

%%%%%%%%%%%%%%%%%%%%%%%%%%%%%%%%%%%%%%%%%%%%%%%%%%%%%%%%%%%%%%%%
\begin{figure}[h]
\begin{center}
\leavevmode
\epsfysize=20truecm \vbox{\epsfbox{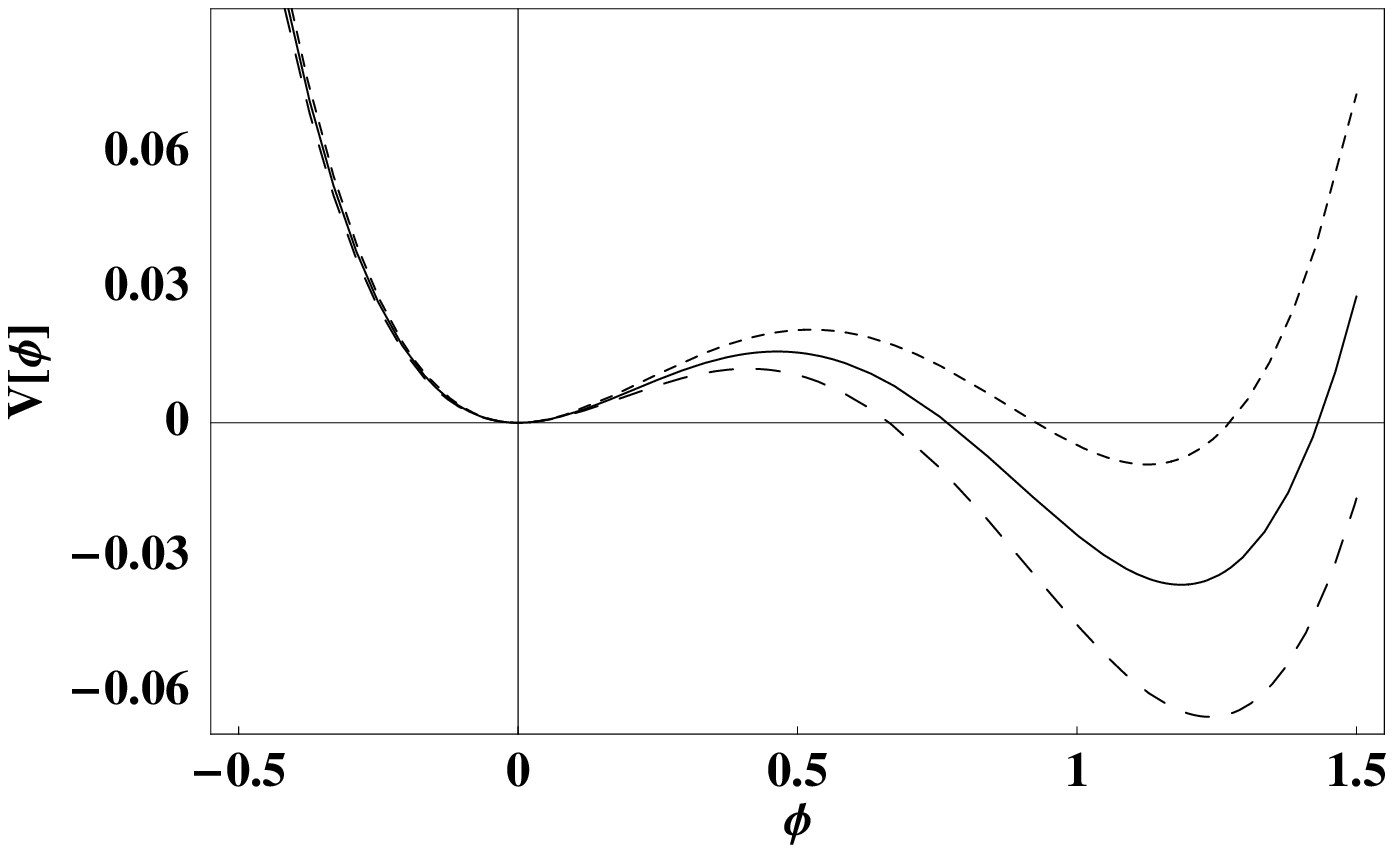}}
\end{center}
\caption{The plot of free energy undergoing periodic variations, with 
$T$ remaining below $T_c$. Order parameter
$\phi$ will be frequently driven via localized resonant oscillations, 
over the barrier into the metastable phase. Note that we use suitably 
scaled, dimensionless variables as explained in Eq.(1).}
\label{Fig.1}
\end{figure}
%%%%%%%%%%%%%%%%%%%%%%%%%%%%%%%%%%%%%%%%%%%%%%%%%%%%%%%%%%%%%%%%%%

 As we will see below, the resulting distribution of the order
parameter is similar to that 
corresponding to the mixed phase which (at low temperatures) exists only 
for short time during the usual first order phase transitions. Here this 
phase is sustained for as long as the free energy keeps varying 
periodically.

 The physical implications of the possibility of achieving such a 
phase can be important. For example, in the context of cosmology, with
oscillating parameter being a background field (as in models of
pre-heating after inflation), this will mean a prolonged
existence of the mixed phase during a first order phase transition,
thereby affecting the final reheating temperature. In the context of
condensed matter systems, if one could achieve such a phase for
water (or any liquid), 
then at temperatures much below the freezing temperature, with the system 
subjected to periodically varying temperature, or more appropriately, 
pressure (density waves with suitable frequency and 
amplitude), microscopic ice crystals may form,
but they will not be able to grow bigger. The whole system will remain
in a mixed phase of microscopic ice crystals continuously forming,
and decaying, but never completing the transition to the frozen phase.
Crucial point is that all this can be achieved while the temperature of
the system remains below the critical temperature.  In 
general, for any first order transition, one may be able to indefinitely 
sustain mixed phase in supercooled (or superheated) state in this
manner. In this context we mention that it has been discussed in the
literature that the structure of supercooled water is very similar to 
the known structure of high density amorphous ice \cite{hda}.
There have also been several studies where
systems in a non-equilibrium state approach quasi-stationary phases
with very large relaxation time scales. For example, recent studies
have discussed similarities between the dynamics of granular materials 
and glassy materials under the influence of external periodic influence 
(see, e.g. \cite{grn}). The existence of (quasi) stationary phases
in such systems \cite{grn,gls} (for relevant time scales) may thus 
appear similar to what we have discussed here, though the underlying 
physics is quite different.  

Another important issue which has been discussed in 
earlier studies of non-equilibrium systems with slow dynamics relates 
to the notion of an effective  temperature \cite{neqT} (see, also,
refs.\cite{grn,gls}). It has been discussed that the 
fluctuation-dissipation relation can be generalized
for such systems with an effective temperature which acts like the 
thermodynamic temperature in the sense that it controls the direction
of heat flow and acts as a criterion for thermalization. In the context
of our model also one would like to know if such an effective temperature
can be defined and whether that effective temperature is significantly
higher than the temperature used in the expression for the free energy
of the system. Thus when we make statements like the
temperature of the system being much below the
critical temperature, we are ignoring, for the purpose of the present
work, these issues relating to the effective temperature etc. for the
non-equilibrium system at hand. We hope to address these issues
in a future work where we will also investigate the issue of
appropriate meaning of re-heat temperature in models of pre-heating 
after inflation.
 
 The paper is organized in the following manner.
In section II we describe our model, and discuss basic physics
involved. Section III presents results of our numerical
simulations. Conclusions and discussions are presented in section IV.

\section{Description of the system}

 We will discuss the dynamics of a system in 2+1 dimensions, described 
by the following free energy (effective potential), which is expressed in 
terms of scaled, dimensionless variables for simplicity of 
presentation \cite{rsntd}.

\begin{equation}
V(\phi) = \frac{1}{4}\phi^2 (\phi - 1)^2 - {\epsilon \over 2} \phi^3
+  {1 \over 4} T(t) \phi^2,
\end{equation}

\noindent with $T(t)$ being the instantaneous temperature of the 
system, which is taken to be spatially uniform. We take the time 
dependence of the temperature to be,

\begin{equation}
T(t) = T_{0} + T_{a}\sin(\omega t).
\end{equation}

 Above,  $\phi$ is a scalar order parameter, and we take $\epsilon = 
0.1$. From Eq.(1), we get the critical temperature $T_c = (1 + \epsilon)^2
- 1 = 0.21$. We discuss the 2+1 dimensional case due to computer 
limitations. We expect similar results for 3+1 dimensions, as the
basic physics of resonant field oscillations remains the same. 
We hope to discuss the 3+1 dimensional case in a future presentation. 
We emphasize that the crucial physics of our model resides in the time 
dependence of the free energy. We have characterized this in terms of a 
time dependent temperature. One could also 
do this by periodically varying some other parameter such as pressure, 
which may be experimentally more feasible, or possibly, even a 
time-dependent external electric or magnetic field (say, for liquid 
crystals). In the context of cosmology this time dependence can be 
achieved by coupling to a background oscillating field, as discussed
below. 

 We use the following equations for field evolution.

\begin{equation}
\partial^2 \phi/\partial t^2 + \eta \partial\phi/\partial t - 
\bigtriangledown^2 \phi + V^\prime(\phi) = 0 .
\end{equation}
 
 Here $\eta$ is the dissipation coefficient. We have included the term with
second order time derivative, (the {\it inertial} term \cite{inrt}), since 
the short time scale dynamics of $\phi$ is of crucial importance here.
We do not include a noise term in Eq.(3). The basic physics we discuss does 
not depend on noise. Also, as discussed above, time dependence of the free
energy could arise from some other source, with temperature kept low to 
suppress any thermal fluctuations (though, we again mention that we are
ignoring here the possibility of defining an effective temperature
\cite{neqT}). In any case, noise due to thermal fluctuations 
should further help in sustaining the mixed phase (e.g. in ref.\cite{fluc} 
it is shown that presence of noise enhances the resonant oscillations of 
the field, leading to enhanced defect production). In a future work we 
will study the detailed effects of noise in our model, especially in the
overdamped limit, and also explore connections with the well 
studied phenomenon of {\it stochastic resonance} in condensed matter 
systems \cite{stch}. 

 As we indicated above, Eq.(3) with oscillatory T(t) is similar to 
the equation for an oscillating inflaton field coupled to another 
scalar field in the models of post-inflationary reheating in the early
universe\cite{infl,infl1}, with $T(t)$ playing the role of the inflaton field.
The results in the present paper (as well as in
ref.\cite{rsntd}), therefore, have similarities with those models, though
there are crucial differences. For example, it has been suggested in
\cite{infl1} that the completion of the transition after inflation may 
get delayed due to parametric resonance
instabilities. Ref.\cite{infl1} discusses the case of a homogeneous
order parameter field coupled to a background oscillating inflaton
field, and it is shown that due to parametric resonance the (uniform)
order parameter keeps flipping between the stable minima and the
metastable one. As we will see below, the actual physics of resonating
order parameter for a first order transition case is much more
complex. For example, field in localized regions is 
resonantly driven to the metastable vacuum where it can shrink
down due to high free energy cost instead of being resonantly driven
back to the stable vacuum. For the universe,
the transition is eventually completed to the low temperature phase, as
the oscillating inflaton field decays. In contrast, in the condensed
matter case, where the periodic driving of temperature/pressure can be 
maintained indefinitely, the continued {\it localized} hopping of 
$\phi$ between the metastable vacuum and the stable vacuum keeps 
the system indefinitely in a mixed phase.

\section{Results}

 We first discuss the situation with fixed $T = T_0 = 0.1$. For $T < T_c$,
$V(\phi)$ has a local minimum at $\phi = 0$ (metastable phase),while the true 
minima (stable phase) occurs at $\phi \equiv \phi_s$ where,

\begin{equation}
\phi_s = \frac{1}{4}[3(\epsilon + 1) + \sqrt{9(\epsilon + 1)^2 - 
8(T(t) + 1)}~~]
\end{equation}

 With $T = 0.1$ we get $\phi_s = 1.186$. The plot of $V(\phi)$ 
at $T = 0.1$ is shown by the solid curve in 
Fig.1. We start with a region of space trapped in the metastable phase with 
$\phi = 0$. At finite temperature, the phase transition takes place by 
nucleation of bubbles of stable phase 
via thermal fluctuations \cite{bbl,bbl0}. (For very low, or zero 
temperature, one may need to consider quantum nucleation of bubbles.)
We simulate the nucleation of these bubbles using techniques developed in 
\cite{bbl}. Bubbles nucleate with (slightly larger than) critical size and 
expand, ultimately filling up the space. The critical bubble profile
is obtained by solving the following field equation \cite{bbl,bbl0}

\begin{equation}
{d^2 \phi \over dr^2} + {1 \over r} {d \phi \over dr} -
V^\prime(\phi) = 0 ,
\end{equation}

\noindent subject to the boundary conditions $\phi(\infty)=0$ and
$d\phi/dr=0$ at $r=0$; where $r$ is the radial coordinate. 
Note that, for our present discussion, the only 
relevant thing is the late time field configuration, and its evolution. 
The bubble configuration obtained by above method provides an adequate 
starting configuration here. Bubble nucleation is achieved by replacing a 
region of the metastable phase (false vacuum) by the bubble profile,
which is suitably truncated with due care of smoothness
of the configuration on the lattice. Subsequent evolution of the 
initial field configuration is governed by Eq.(3) with ${\dot \phi} = 
0 $ at $t = 0$.

%%%%%%%%%%%%%%%%%%%%%%%%%%%%%%%%%%%%%%%%%%%%%%%%%%%%%%%%%%%%%%%%
%\vskip -0.25in
\begin{figure}[h]
\begin{center}
\leavevmode
\epsfysize=20truecm \vbox{\epsfbox{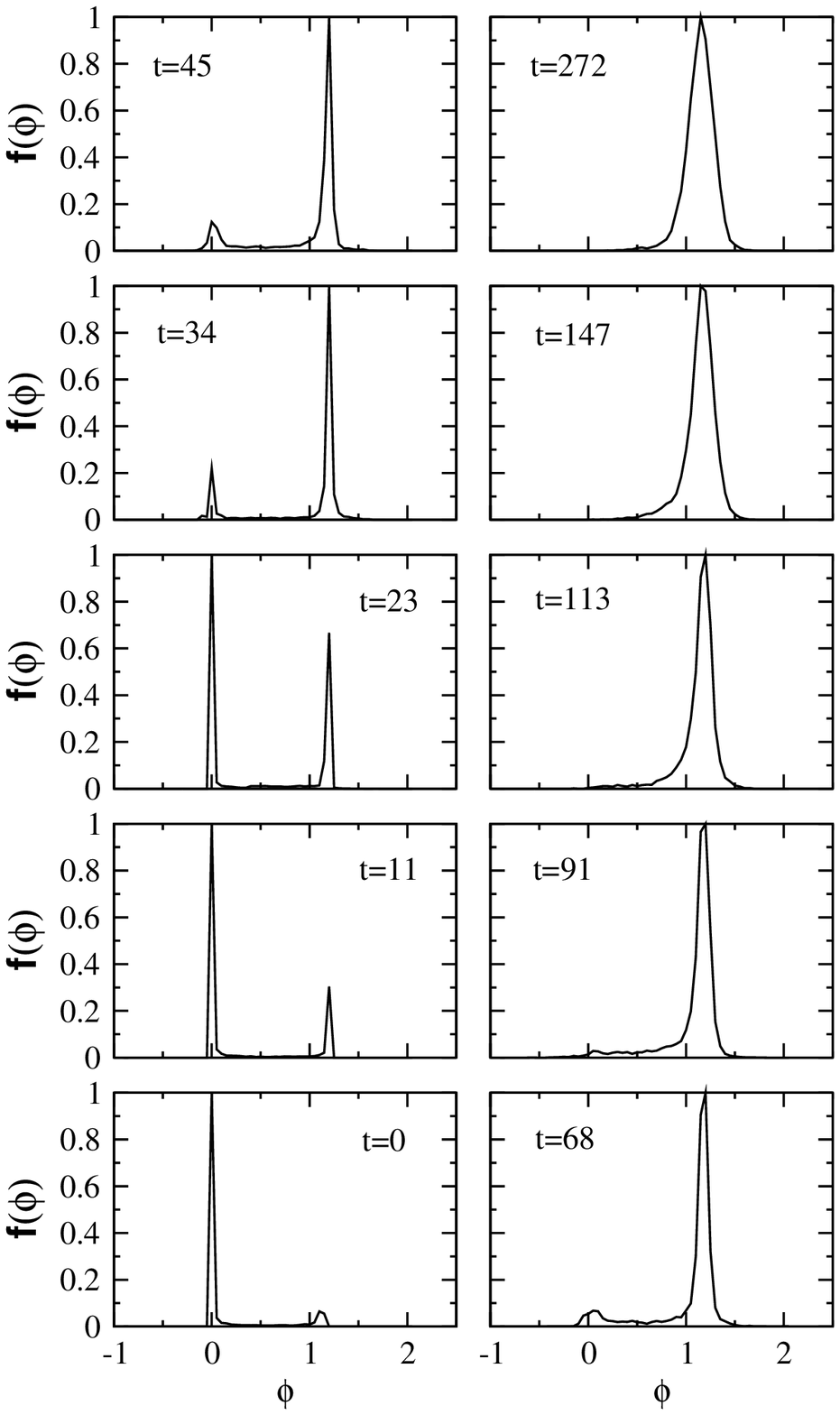}}
\end{center}
\caption{The plots of the distribution $f(\phi)$ vs. $\phi$ for the 
phase transition process simulated via bubble nucleation with $\omega = 0$
in Eq.(2). Plots showing presence of two peaks correspond
to the mixed phase. At late stages, this mixed phase completely
disappears, and one gets a pronounced peak at the stable value of
$\phi = \phi_s$ indicating the completion of the phase transition.}
\label{Fig.2}
\end{figure}
%%%%%%%%%%%%%%%%%%%%%%%%%%%%%%%%%%%%%%%%%%%%%%%%%%%%%%%%%%%%%%%%%%

 Nucleation of  several bubbles is achieved by randomly choosing the 
location of the center of each bubble with some specified probability
per unit volume \cite{bbl}. (For simplicity we nucleate all the bubbles
at the initial time. Time depended nucleation will not affect the 
phenomenon discussed in this paper.) If there is an overlap with a 
previously nucleated bubble, then nucleation of the new bubble is skipped.  
The simulation is done on a square lattice with periodic boundary condition, 
i.e on a torus. The field configuration is evolved by using a discretized 
version of Eq.(3), using a second order, staggered, leapfrog algorithm. The 
size of the lattice taken is $900 \times 900$  with $\Delta x = 0.16$ 
and $\Delta t \simeq 0.0075$. Simulations were 
carried out on HP workstations at the Institute of Physics, Bhubaneswar.
As we are interested in the nature of the mixed phase, we present our
results in terms of plots of the fraction $f(\phi_0)$ representing the 
fractional volume of the region (calculated by counting the number of 
lattice sites) where $\phi$ lies between $\phi_0$ and 
$\phi_0 + \Delta \phi$, for a suitable choice of $\Delta \phi$. We use 
$\Delta \phi = 0.05 \simeq 0.042 \phi_s$. Here $\phi_s \simeq 1.186$ for $T = 
0.1$. The plots are normalized by taking the largest value of $f(\phi) = 1$
for the given plot. 

 Fig.2 shows the results of the simulation (with $\eta = 0$). The plot at $t 
= 0$ shows the initial $f(\phi)$ which has a sharp peak at the background
metastable phase, $\phi = 0$, and a smaller peak at $\phi = \phi_s \simeq 
1.19$ corresponding to stable phase inside the nucleated bubbles. (In
each of the cases discussed here and below, 3 bubbles were nucleated.) The
relative heights of the two peaks changes as the bubbles expand. At 
sufficiently late times, the only peak remains at $\phi = \phi_s$, 
indicating the completion of the phase transition to the stable phase. 
Plot of $f(\phi)$ remains essentially unchanged for all subsequent times. 
Bubble wall coalescence leads to $\phi$ oscillations which persist, due to 
absence of dissipation, contributing to the width of the peak at $\phi_s$ 
at late times. The nature of the plots will remain the same in the case 
even with non-zero $\eta$, which only affects the bubble wall velocities, 
and damping of $\phi$ oscillations (leading to sharper peak at late times).

  We now repeat this simulation with non-zero value of $\omega$ in Eq.(2). 
We use $T_0 = 0.1$ and $T_a = 0.08$, and $\omega = 0.78$. With $T_c = 
0.21$, we note that the maximum value of $T$ remains
below $0.9 T_c$, while the average $T$ is below $0.5 T_c$. Upper and lower
dashed curves in Fig.1 show the plots of $V(\phi)$ at $T = T_0+T_a$, and  
$T = T_0-T_a$, respectively. We use $V(\phi)$ in Eq.(5) with 
$T = T_0 + T_a = 0.18$ for calculating the bubble profile.
This is to make sure that the bubbles continue
to expand during the oscillation of $T$. Size of the critical
bubble obtained from Eq.(5) is about 24.1.  
$T$ in Eq.(1) is taken to be spatially uniform, with its 
periodic variation given by Eq.(2). The 
choice of frequency $\omega$ was guided by the range of frequency 
required to induce resonance for the case of spatially uniform field
evolved by Eq.(3) (as in ref.\cite{rsntd}). 
(See,also, in this context, the discussion relating to the parametric
resonance in ref.\cite{ll}.) We find that resonance 
happens when  $\omega$ lies in a certain range. We are assuming
that for the relevant range of $\omega$  here, the system can be
considered in quasi-equilibrium so that the use of time
dependent $V$ makes sense. This frequency range, for 
which resonant oscillations of $\phi$ occurs, depends on $T_{0}$ as 
well as on $T_a$, with the range becoming larger as $T_{0}$ approaches $T_c$. 
Basically, the value of $\omega$ should be such that $\phi$ oscillations
should be affected significantly by the changes in the shape of $V(\phi)$
(as argued in ref.\cite{rsntd}). Thus one expects that
$\omega$ should be roughly of the same order as the 
natural frequency of oscillation of the order parameter near the stable 
minimum of the free energy. With the above values of $T_0$ and $T_a$, the 
range of $\omega$ is found to be between 0.68 and 0.98. We present 
results for $\omega = 0.78$.

%%%%%%%%%%%%%%%%%%%%%%%%%%%%%%%%%%%%%%%%%%%%%%%%%%%%%%%%%%%%%%%%
%\vskip -0.25in
\begin{figure}[h]
\begin{center}
\leavevmode
\epsfysize=20truecm \vbox{\epsfbox{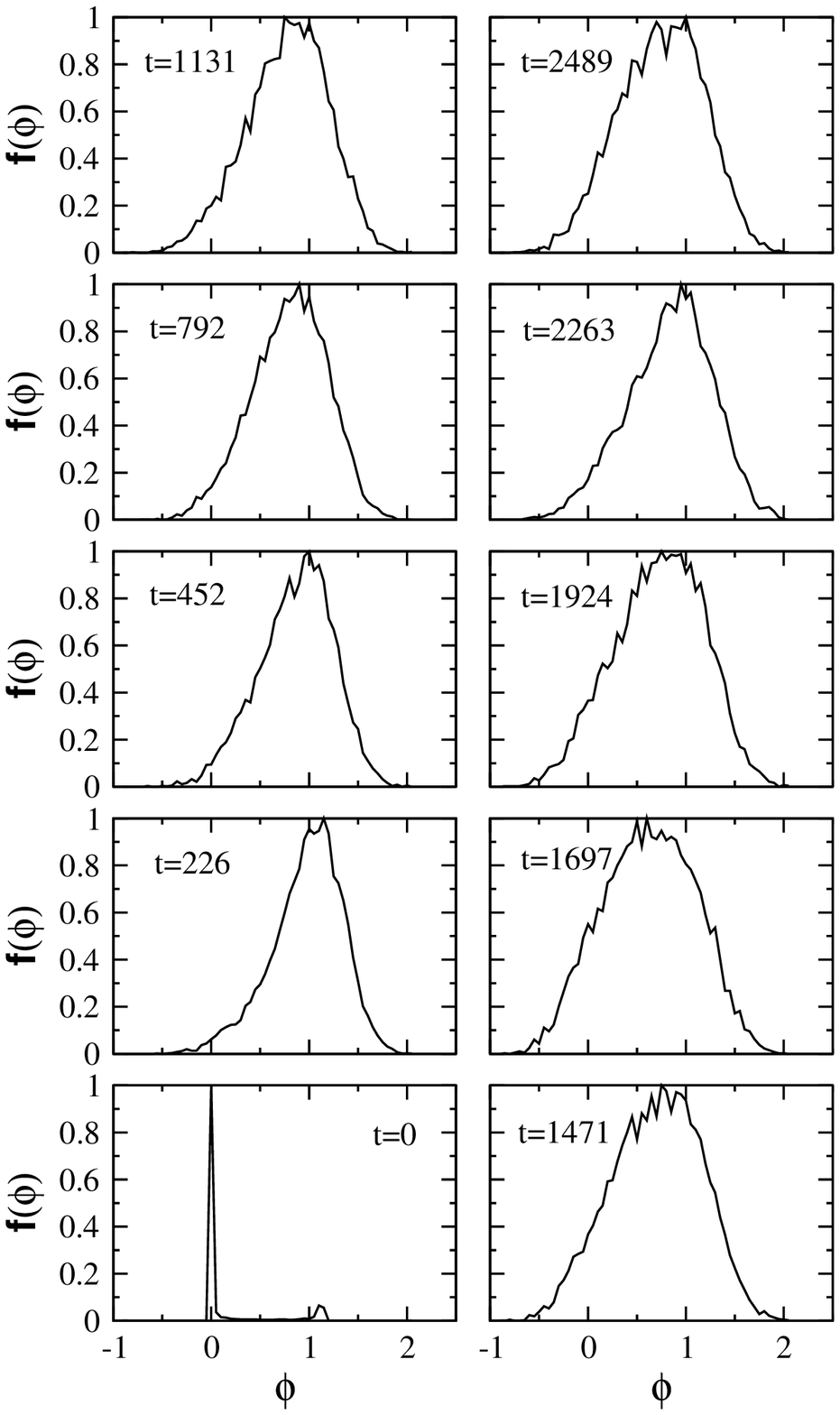}}
\end{center}
\caption{The plots of the distribution $f(\phi)$ vs. $\phi$ for
the case with $\omega = 0.78$. Plots for late times
show significant support at small values of $\phi$. $f(\phi=0)$ 
keeps fluctuating between 20 \% to 50 \% at all late times, 
clearly demonstrating that the transition never gets completed to the
stable phase.}
\label{Fig.3}
\end{figure}
%%%%%%%%%%%%%%%%%%%%%%%%%%%%%%%%%%%%%%%%%%%%%%%%%%%%%%%%%%%%%%%%%%

As discussed above, this periodic variation in $V(\phi)$ drives $\phi$ 
periodically and leads to frequent resonances \cite{rsntd}. Occasionally, 
$\phi$ in a small region is able to get resonantly driven all the way 
over the barrier. Thus, in the regions where $\phi \simeq \phi_s$, 
occasionally a small patch of metastable value $\phi = 0$ forms. This patch 
is energetically unstable, so it shrinks. Or, it could also oscillate to
the stable phase. Crucial point here is that even as bubbles all
coalesce, $\phi$ still keeps flipping over the barrier to the
metastable value $\phi = 0$ keeping the system in a mixed phase.
This will keep happening as long as the system is periodically
driven with appropriate value of $\omega$ in Eq.(2). We mention that 
even the largest patch we find, where $\phi$ flips over the barrier 
to the metastable value, has size much smaller than the critical
bubble size, otherwise one could nucleate stable phase bubbles there. 

 Fig.3 shows the evolution of the distribution $f(\phi)$ for this case. 
The plot at $t = 0$ is the same as in Fig.2. However, now, as bubbles 
coalesce, instead of getting a more and more pronounced peak only at the 
stable value of $\phi = \phi_s$, we are getting a very broad distribution 
of $f(\phi)$. Note, in particular, that $f(\phi)$ at the metastable value 
$\phi = 0$ is never zero, in complete contrast to the situation in Fig.2.

%%%%%%%%%%%%%%%%%%%%%%%%%%%%%%%%%%%%%%%%%%%%%%%%%%%%%%%%%%%%%%%%
\begin{figure}[h]
\begin{center}
\leavevmode
\epsfysize=14truecm \vbox{\epsfbox{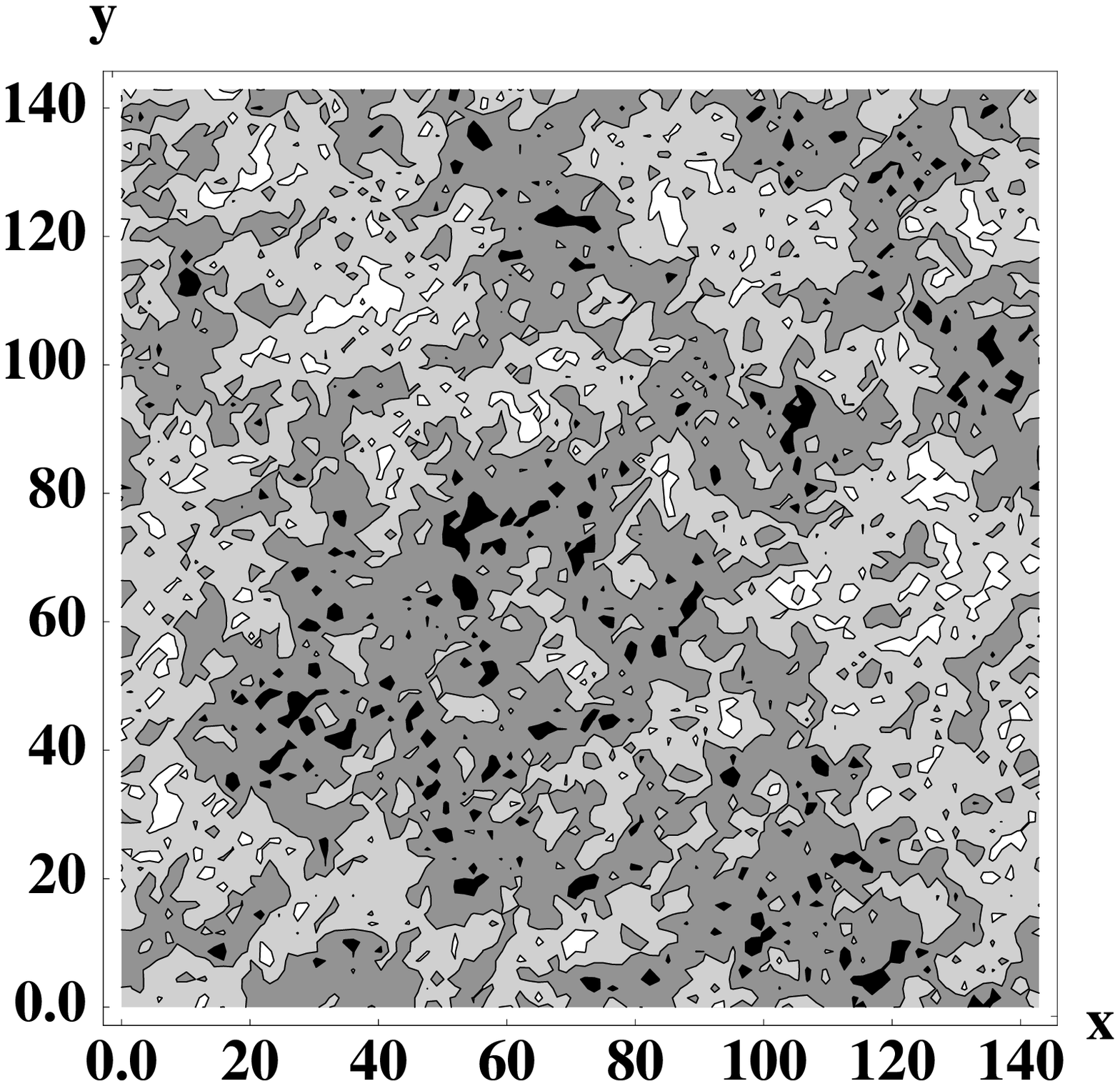}}
\end{center}
%\vskip -1.6in
\caption{Contour plot of spatial distribution of $\phi$ at $t = 1697$. 
White regions correspond to $\phi = 0$, while black regions correspond to 
the largest value of $\phi$ at this stage. Mixing of dark and light 
regions is a clear evidence of the existence of the mixed phase.}
\label{Fig.4}
\end{figure}
%%%%%%%%%%%%%%%%%%%%%%%%%%%%%%%%%%%%%%%%%%%%%%%%%%%%%%%%%%%%%%%%%%

 In fact the ratio of the volume where $\phi = 0$, to the volume where 
$\phi$ is near the oscillating stable vacuum  (where $f(\phi) \simeq
1$  with our normalization) 
keeps fluctuating between about 20\% to almost 50 \% at all late times.
(Non-zero values of $f(\phi)$ at negative $\phi$ result due to oscillations
of $\phi$ about $\phi = 0$.)
This is the direct evidence that the system never completes the phase 
transition to the stable phase, and remains in the mixed phase
indefinitely. This is further evidenced by the contour plot of $\phi$ in 
Fig.4. Here, shading represents the value of $\phi$, with white region
corresponding to the smallest value of $\phi$ (which is negative due to
$\phi$ oscillations about $\phi = 0$), while black region corresponding to 
the largest value of $\phi$ in a given plot. Existence of mixed phase is
clear by the mixing of dark and light regions. Another
important thing to note is that neither white, nor black regions
occur in large sizes (compared to the critical bubble radius $\simeq 24.1$
as mentioned above). Further, value of $\phi$ keeps fluctuating rapidly
at any given point, as we will see below. Thus there are no regions 
where even in an approximate sense phase transition can be said to be 
completed to the stable phase. Similar features can be seen in Fig.5 which
shows 3-dimensional plot of $\phi$ in a small portion of the lattice. 
Fluctuations in $\phi$ are seen to fill up the entire physical region.

%%%%%%%%%%%%%%%%%%%%%%%%%%%%%%%%%%%%%%%%%%%%%%%%%%%%%%%%%%%%%%%%
%\vskip -0.25in
\begin{figure}[h]
\begin{center}
\leavevmode
\epsfysize=18truecm \vbox{\epsfbox{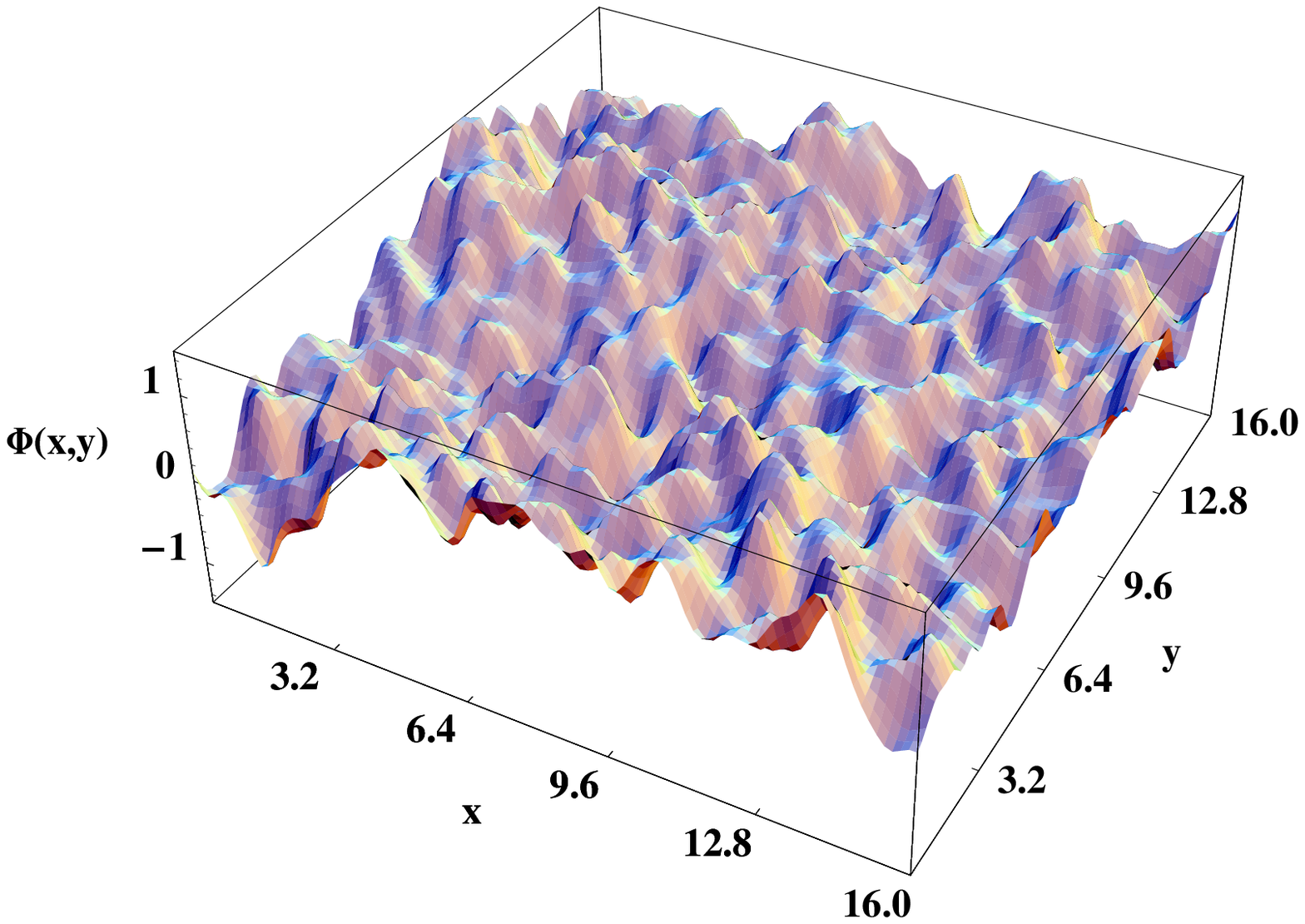}}
\end{center}
%\vskip -1.6in
\caption{3-dimensional plot of $\phi$ at $t = 2602$ for a small portion of
the lattice.}
\label{Fig.5}
\end{figure}
%%%%%%%%%%%%%%%%%%%%%%%%%%%%%%%%%%%%%%%%%%%%%%%%%%%%%%%%%%%%%%%%%%

  To better understand the resonantly driven dynamics of field here, we 
show, in Fig.6, plot of $\phi(x)$ at an arbitrary fixed point for a small
time duration. While oscillating,
$\phi$ bounces back at some point $\phi > \phi_s$ ($ \simeq$ 1.2 for
the $V(\phi)$ shown by the solid plot in Fig.1), and towards the
metastable minimum, at some negative value of $\phi$.  That is, resonant
oscillations drive field from one minimum of $V(\phi)$ to the other. 
In between, $\phi$ executes oscillations about $\phi = 0$, and about
the stable vacuum. The detailed dynamics of $\phi(x)$ at a given point
is much more complex than simple oscillations about the two minima 
(and resonant driving in between the two), due to the effect of
density waves coming from other regions, shrinking of regions of
metastable vacuum etc. (As we mentioned above, such features, which
strongly affect the nature of resulting mixed phase, could not be
seen in ref.\cite{infl1} where order parameter was taken to be
uniform). Still, the plot in Fig.6 shows that the qualitative
aspects of the field dynamics are dominated by oscillatory motions,
and not by some random noise (which would be the case for thermal
fluctuation dominated dynamics).  This is consistent with the physical 
picture outlined above that the system is in a 
complete non-equilibrium state, being continuously driven by the
periodically varying temperature in Eq.(1).

%%%%%%%%%%%%%%%%%%%%%%%%%%%%%%%%%%%%%%%%%%%%%%%%%%%%%%%%%%%%%%%%
%\vskip -0.25in
\begin{figure}[h]
\begin{center}
\leavevmode
\epsfysize=15truecm \vbox{\epsfbox{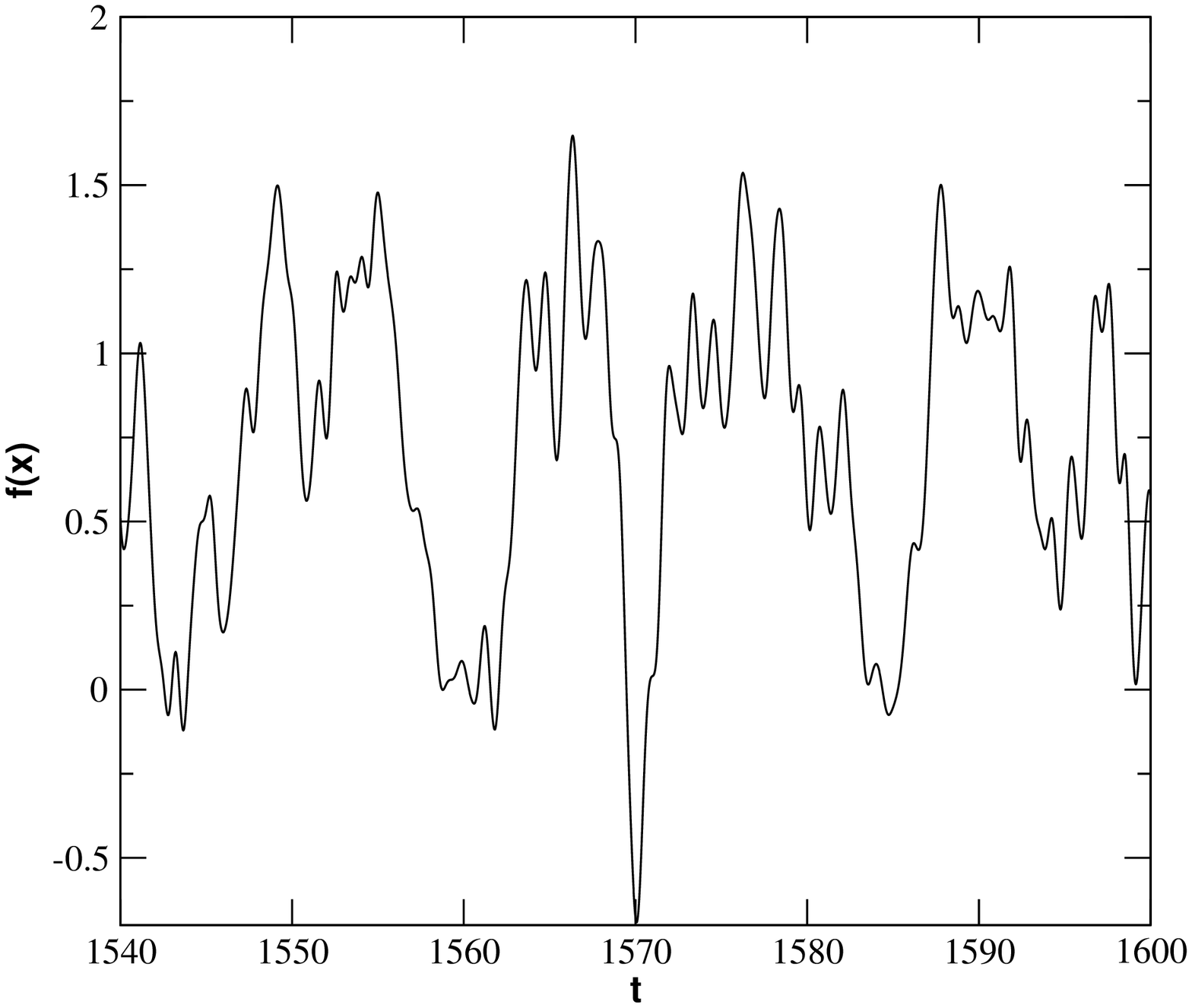}}
\end{center}
\caption{Evolution of $\phi(x)$ at an arbitrary, fixed point. The
detailed dynamics of $\phi(x)$ at a given point is more complicated 
than simple oscillations about the two minima (and resonant driving
in between the two), due to the effect of density waves coming from 
other regions, etc.} 
\label{Fig.6}
\end{figure}
%%%%%%%%%%%%%%%%%%%%%%%%%%%%%%%%%%%%%%%%%%%%%%%%%%%%%%%%%%%%%%%%%%

 We now consider the case of non-zero dissipation. As expected, for 
large values of $\eta$, $\phi$ oscillations damp fast,
suppressing resonant oscillations (as found in ref. \cite{rsntd}).
For small $\eta$, as in Fig.7, oscillations are somewhat suppressed, but
the  mixed phase is still sustained. In Fig.7 we give plots of 
$f(\phi)$ for the case with $\eta = 0.01$. As mentioned above, in view of
results in ref.\cite{fluc}, we expect that inclusion of noise term in 
Eq.(3) will lead to enhanced resonances (possibly even in the
overdamped limit). 
Note that the distribution of $f(\phi)$ for the mixed phase in Fig.3,7
is very different from the mixed phase plots in Fig.2. It will be 
interesting to explore if there are essential physical differences
between these two types of mixed phases. It is possible that with the
inclusion of thermal fluctuations this difference may reduce.

%%%%%%%%%%%%%%%%%%%%%%%%%%%%%%%%%%%%%%%%%%%%%%%%%%%%%%%%%%%%%%%%
\begin{figure}[h]
\begin{center}
\leavevmode
\epsfysize=20truecm \vbox{\epsfbox{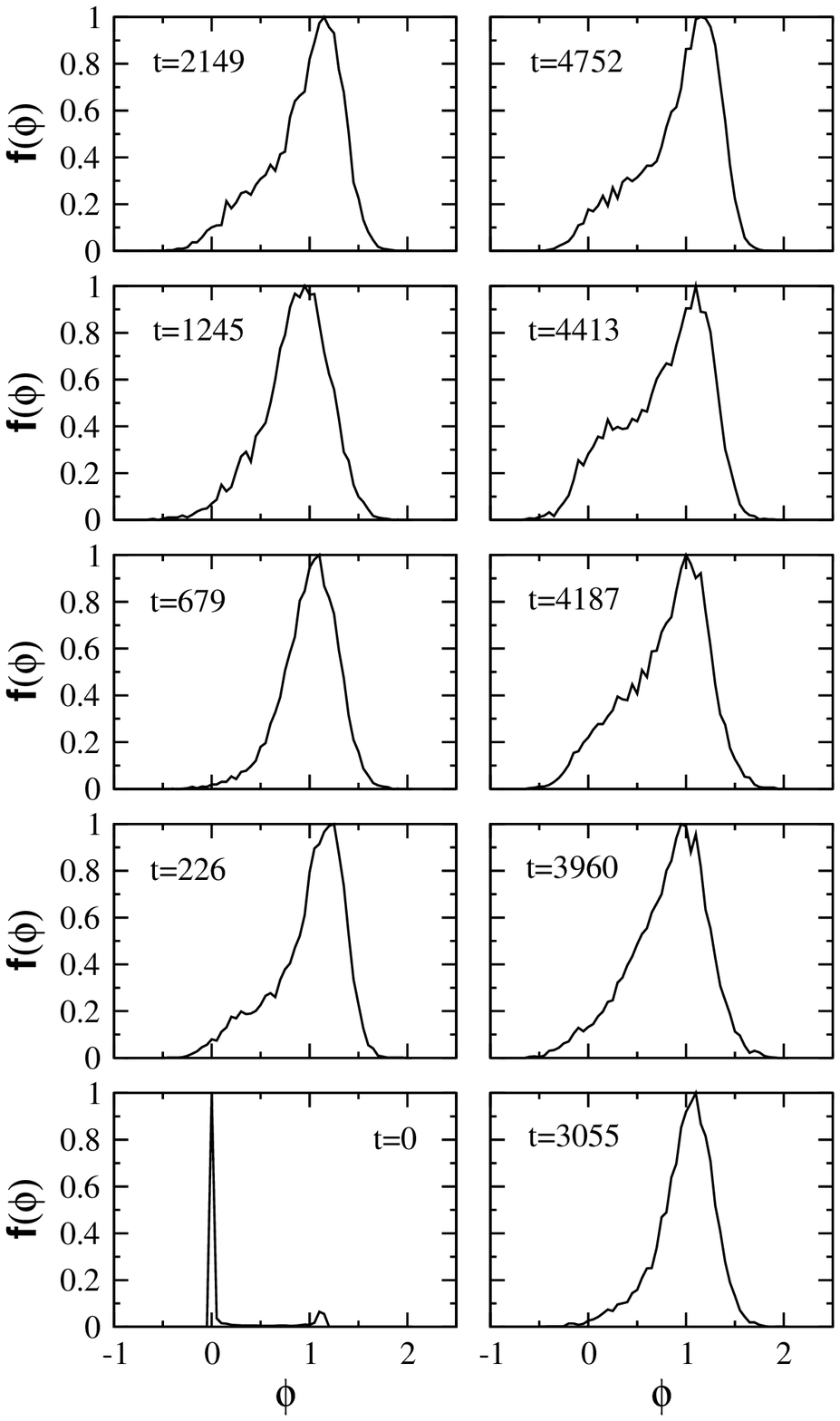}}
\end{center}
\caption{The plots of the distribution $f(\phi)$ vs. $\phi$ for
the case with $\eta = 0.01$. Even here, we find that plots for late 
times show significant support at small values of $\phi$. $f(\phi=0)$ 
keeps fluctuating between 5 \% to 20 \% at all late times.}
\label{Fig.7}
\end{figure}
%%%%%%%%%%%%%%%%%%%%%%%%%%%%%%%%%%%%%%%%%%%%%%%%%%%%%%%%%%%%%%%

  It is important to mention here that the existence of a local minimum
(at $\phi = 0$ in Eq.(1)) is crucial in getting significant support for 
$f(\phi)$ near $\phi = 0$. When $\phi$ is resonantly driven over the barrier 
to that local minimum, it may get trapped there, oscillating about
$\phi = 0$,  until the region shrinks
away (as mentioned above, this region is always too small to nucleate
another bubble there). When we use a free energy similar to that in
ref.\cite{rsntd} (appropriate for a second order transition), 
we find that distribution of $f(\phi)$ still broadens
at $\phi_s$, but it does not develop significant support at
$\phi = 0$. (That is the situation with complex 
$\phi$. If we take real $\phi$ with the $V$ in ref.\cite{rsntd}, 
then significant volume fraction with $\phi$ = 0 arises due to regular 
production and shrinking of extended domain wall defects.) However, we 
emphasize that even in the case of ref.\cite{rsntd}, the distribution
of $\phi$ at late times resembles the situation of an equilibrium
system which is close to the transition point, even when the
temperature remains much below the critical values. 
(We make this statement in a loose sense, basically focusing
on qualitative features of domain structure and fluctuations).
The physical properties of that system
will therefore be completely different from the one expected from
the system at the value of the temperature used in $V$, just as
demonstrated here for the case of first order transition. 

\section{Discussion and Conclusions}

  Our results show a very interesting possibility, and at the same time 
raise many questions. The physical nature of such a system looks very
different from the system at the temperature used in $V$ in Eq.(1). 
Though we are discussing the situation of non-equilibrium, the system 
seems to reach a state of quasi-equilibrium, or, more appropriately,
a stationary state, where its average properties do not fluctuate too 
much, such as the distribution of fraction $f(\phi)$ in Figs.3,7.
We mention here again that there may be  a possibility of defining
some sort of effective temperature here, following the discussions 
in the literature \cite{neqT}. It will then be interesting to 
determine how the properties of the system we observe relate to the
expected properties of the system at that effective temperature.

 Also, as we discussed earlier, though we have characterized the 
periodically varying term in $V(\phi)$ in Eq.(1) in terms of a periodically
varying temperature, it could be achieved in various different ways.
For example, in cosmology, coupling to a background oscillating 
inflaton field will give rise to required driving of the order
parameter field \cite{infl,infl1}. In many condensed matter systems 
(e.g. liquids) periodic variation of pressure may be a better choice
as that can easily be induced by density waves. (From that point of 
view it will be interesting to consider periodic temporal as well as 
spatial variations of the parameter in $V(\phi)$.)
For some systems, such as liquid
crystals, periodically varying external fields (electromagnetic field)
may be more suitable. We emphasize that, though actual numbers (e.g.
fraction of the system in the metastable phase etc.) will vary from
one system to other, (and from $2+1$ dimensional case discussed here to 
the $3+1$ dimensional case), the basic physics of the phenomenon we have 
discussed here appears very robust. A rapidly oscillating free energy, due 
to periodic variation of its parameter(s), will in general be expected to
lead to resonant oscillations of the order parameter, with appropriate
frequency and amplitude of the oscillating parameter. This will lead
to periodic, localized creations of regions with metastable phase, even
when the temperature (and other parameters) of the system always remain
much below the transition point (or above it, for the case of superheating). 
Resulting phase is similar to the mixed phase which can be sustained for 
as long as the system is periodically driven.

 For cosmology, this phenomenon will imply an extended period of
mixed phase, until the background oscillating inflaton field decays. 
This will affect the value of final reheat temperature after
inflation. In the context of condensed matter systems this
possibility of indefinitely sustaining mixed phase of systems at 
temperatures  below the critical temperature can have very important 
implications, especially for supercooling liquids. It is tempting 
to speculate that if such a possibility can be realized 
for water, then it would imply that supercooling organisms may 
be possible (when subjected to, say, pressure waves of suitable frequency 
and amplitude) without freezing, as large ice crystals will never form.
(Note, again, we are ignoring here the issues relating to the
effective temperature \cite{neqT} which could be defined
for the non-equilibrium system at hand. It is possible that such
an effective temperature may not be as low as the temperature
which is used in the expression of the free energy in Eq.(1).
We hope to investigate these issues in a future work.)
As mentioned above, the appropriate range of $\omega$, for the phenomenon
we discuss, will be expected to be of the  order of the natural frequency
of oscillation of the order parameter near the stable minimum of the
free energy of the specific system being considered. Thus, it should be 
possible to experimentally verify this possibility in condensed matter
systems by subjecting liquids (water) to density waves using a 
varying range of frequency and amplitude, while supercooling below $T_c$. 

\vskip .2in
\centerline {\bf ACKNOWLEDGEMENTS}
\vskip .1in

   We are very thankful to A.P. Balachandran, Sanatan Digal, Biswanath Layek,
Rashmita Sahu, and Supratim Sengupta for useful discussions and comments. 
AMS would like to acknowledge the hospitality of the Physics Dept. 
Univ. of California, Santa Barbara while the paper was being written. 
His work at UCSB was supported by NSF Grant No. PHY-0098395.

%%%%%%%%%%%%%%%%%%% 

%\end{multicols}
\end{document}